\documentclass[onecolumn]{article}
\usepackage[letterpaper, margin=2cm]{geometry}

\usepackage[T1]{fontenc}
\usepackage{graphicx}
\usepackage{xcolor}
\usepackage{amsmath}
\usepackage{booktabs}
\usepackage[colorlinks=true, linkcolor=blue, citecolor=red, urlcolor=magenta]{hyperref}

\usepackage{authblk}

\begin{document}

\title{An Autonomy Loop for Dynamic HPC Job Time Limit Adjustment}

\author[1]{Thomas Jakobsche}
\author[1]{Osman Seckin Simsek}
\author[2]{Jim Brandt}
\author[2]{Ann Gentile}
\author[1]{Florina M. Ciorba}

\affil[1]{University of Basel, Switzerland\\

\texttt{\{thomas.jakobsche,osman.simsek,florina.ciorba\}@unibas.ch}}

\affil[2]{Sandia National Laboratories, CA and NM, US\\

\texttt{\{brandt,gentile\}@sandia.gov}}

\date{}

\maketitle

\begin{abstract}
    High Performance Computing (HPC) systems rely on fixed user-provided estimates of job time limits. 
    These estimates are often inaccurate, resulting in inefficient resource use and the loss of unsaved work if a job times out shortly before reaching its next checkpoint. 
    This work proposes a novel feedback-driven autonomy loop that dynamically adjusts HPC job time limits based on checkpoint progress reported by applications. 
    Our approach monitors checkpoint intervals and queued jobs, enabling informed decisions to either early cancel a job after its last completed checkpoint or extend the time limit sufficiently to accommodate the next checkpoint.
    The objective is to minimize \textit{tail waste}, that is, the computation that occurs between the last checkpoint and the termination of a job, which is not saved and hence wasted.
    Through experiments conducted on a subset of a production workload trace, we show a 95\% reduction of \textit{tail waste}, which equates to saving approximately 1.3\% of the total CPU time that would otherwise be wasted.
    We propose various policies that combine early cancellation and time limit extension, achieving \textit{tail waste} reduction while improving scheduling metrics such as weighted average job wait time.
    This work contributes an autonomy loop for improved scheduling in HPC environments, where system job schedulers and applications collaborate to significantly reduce resource waste and improve scheduling performance.
\end{abstract}

\section{Introduction} \label{section:introduction}

\textit{Monitoring} in HPC systems is a well-explored domain that generates large volumes of data.
Typically, monitoring involves systematic collection of metrics and generating alerts when thresholds are exceeded.
\textit{Operational Data Analytics} has traditionally relied on postmortem analysis or manual decisions, which is becoming increasingly unfeasible given the growing scale of systems~\cite{brandt2023driving}.
The move towards \textit{Observability} in HPC acknowledges that simply logging counters is insufficient due to complex and dynamic application behavior.
Application-specific insights are needed while jobs are still running, to enable dynamic diagnosis, proactive understanding of application behavior, and actionable feedback~\cite{yokelson2024soma}.
Autonomy loops in the form of feedback-driven mechanisms that combine monitoring, data analytics, and automated response have recently been proposed to streamline HPC operations~\cite{boito2023autonomy}.
One such approach is to adjust the time limits of running jobs, aiming to reduce wasted compute time due to misaligned timeouts of jobs that use fixed-time interval checkpointing.

HPC systems often rely on fixed user-estimated job time limits that are known to be notoriously inaccurate and represent a persistent challenge in HPC~\cite{bailey2005user,patel2020job,soysal2019quality}.
Traditional job schedulers, such as Slurm, terminate a job if it exceeds its user-estimated time limit.
This leads to wasted resources if the jobs timeout before completion or just before their next checkpoint.
In this work, we specifically focus on reducing \textit{tail waste} (i.e., the unsaved computations made between the last completed checkpoint and the timeout).

Mechanisms like Slurm's \texttt{OverTimeLimit} offer a grace time beyond the time limit~\cite{yoo2003slurm}, but are applied as \textit{blanket} parameters to all jobs, without leveraging insights into the checkpoint progress of applications; this could also mean giving extra time to applications that are stuck and do not make progress.
While dynamic-time interval checkpointing could potentially align checkpoint schedules to time limits (instead of aligning time limits to checkpoint schedules), it is not widely adopted by users~\cite{el2016understanding}.
These shortcomings result in wasted compute time, suboptimal resource usage, and an increased risk of partially lost work when jobs are terminated shortly before the next checkpoint.

We propose a novel mechanism for dynamically adjusting job time limits which is implemented as an autonomy loop.
By monitoring queued jobs and checkpoint intervals reported from applications, this loop either cancels jobs immediately after their last completed checkpoint or extends their time limit to accommodate the next checkpoint. This reduces resource waste while allowing jobs to align time limits with their checkpoints, thus completing gracefully. As a result, we achieve an approximately 95\% reduction of \textit{tail waste} (i.e., the unsaved computations made between the last completed checkpoint and the timeout), saving approximately 1.3\% of total CPU time that would otherwise be wasted. We introduce three policies for different scenarios: \texttt{Early Cancellation}, which cancels a job after its last completed checkpoint that still fits into the initial time limit, \texttt{Time Limit Extension}, which always extends the time limit to allow the next checkpoint, and a \texttt{Hybrid} approach, which extends the time limit only if it does not delay other jobs, otherwise canceling the job early.

We evaluated the proposed solution on a subset of a production workload trace from \texttt{CINECA}'s Marconi supercomputer~\cite{borghesi2023m100}. 
The extracted job configurations were executed as synthetic jobs on a University research cluster (approximately 800 jobs on 20 nodes).
By comparing our proposed approach with the \texttt{Baseline} (no time limit adjustments), we quantify the impact of early cancellations, time limit extensions, and their hybrid combination. \\

This work makes the following contributions:

\begin{enumerate}
    \item \textbf{An autonomy loop for dynamic time limit adjustment}: We introduce a feedback-driven mechanism that continuously monitors checkpoint progress and queued jobs to dynamically adjust job time limits to reduce \textit{tail waste}.
    \item \textbf{Integration with existing schedulers:} The proposed solution seamlessly integrates with Slurm through standard commands, reducing the \textit{barrier to its adoption} in production HPC environments.
    \item \textbf{A flexible policy framework:} We define and evaluate multiple policies (\texttt{Early cancellation}, \texttt{Time limit extension}, and a \texttt{Hybrid} approach) to accommodate different scheduling priorities of system operators.
    \item \textbf{Empirical validation:} We show the effectiveness of our approach by executing jobs from a production trace on a research HPC system, showing quantifiable improvements in reducing wasted compute time.
\end{enumerate}

The remainder of this paper is structured as follows.
Section~\ref{section:background} offers background and positions our work.
Section~\ref{section:methodology} presents the methodology.
Section~\ref{section:evaluation} contains the evaluation.
The results are provided in Section~\ref{section:results} and discussed in Section~\ref{section:discussion}.
Section~\ref{section:related-work} reviews the related work, while Section~\ref{section:conclusion} concludes the work.

\section{Background} \label{section:background}

Fixed-time interval checkpointing is often based on rule-of-thumb intervals or mathematical formulas for optimal checkpoint periods like Young-Daly~\cite{daly2006higher,young1974first}.
Dynamic-time interval checkpointing approaches adjust checkpointing in real-time and include \textit{outside} approaches, such as distributed multithreaded checkpointing (DMTCP), which works without changes to system or applications~\cite{ansel2009dmtcp}.
The vast majority of HPC applications do not use dynamic checkpointing approaches due to perceived complexity by HPC users~\cite{bautista2024survey,el2016understanding}. 
In our context, it is important that the real cost of checkpointing is not just the checkpoint I/O time but also (and often more importantly) the time to re-execute the work lost after a job interruption. 

Recently, a number of datasets from Tier-0 supercomputers have been made public.
M100~\cite{borghesi2023m100} and the PM100 subset~\cite{antici2023pm100,andrea_borghesi_2023_7588815} (which includes job configurations) from \texttt{CINECA}'s Marconi, F-DATA~\cite{antici_2024_11467483} from \texttt{RIKEN}'s Fugaku, and the data~\cite{oedi_5860} from \texttt{NREL}'s Eagle supercomputer.
Our experiments use job traces from the PM100 dataset, but can easily be expanded to other datasets.

Slurm employs two schedulers for job execution: the main scheduler (\texttt{SchedMain}) which prioritizes the launch of higher priority jobs and the backfill scheduler (\texttt{SchedBackfill}) which evaluates all jobs (regardless of priority) and attempts to schedule those that do not delay the start of a higher priority job.

A foundational theoretical framework for self-managing systems was introduced in a seminal work on autonomic computing~\cite{kephart2003vision}, which forms the basis of our use of feedback loops and dynamic adjustment mechanisms.

Our approach is positioned with respect to the concepts of \textit{malleability} and \textit{dynamism} in HPC. 
Malleability refers to modifying the number of resources allocated to a job during execution, such as adding or removing compute nodes. 
In contrast, dynamically adjusting job time limits falls under the broader notion of dynamism rather than malleability~\cite{tarraf2024malleability}.

\section{Methodology} \label{section:methodology}

This section describes the proposed methodology, the time limit adjustment policies, and defines \textit{tail waste} and the scheduling metrics considered in this work.

\begin{figure}[!h]
    \centering
    \includegraphics[width=0.8\textwidth]{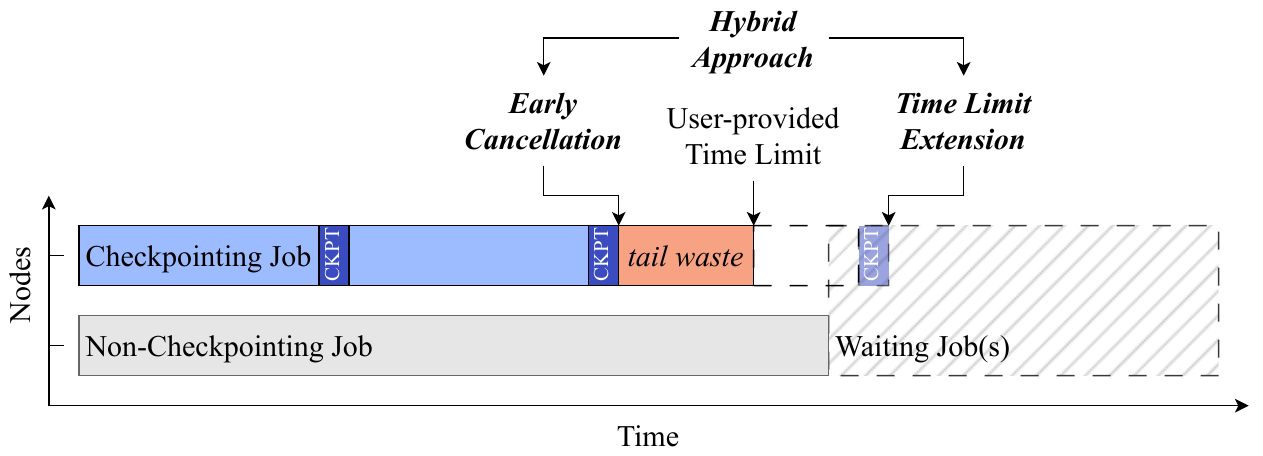}
    \caption{The key mechanism at the core of the autonomy loop.
    For the checkpointing job, a misalignment of its user-provided time limit and checkpointing schedule leads to \textit{tail waste}.
    The non-checkpointing job does not report checkpoint progress and is not further considered by the daemon.
    \textit{Tail waste} is avoided by adjusting the time limit to align with the checkpoint schedule through different policies:
    The \texttt{Early cancellation} policy which cancels the job early after the last successful checkpoint, the \texttt{Time limit extension} policy which extends the time limit to the next checkpoint, even if this delays other jobs, and the \texttt{Hybrid} approach policy which either extends the job time limit if this does not delay other jobs or cancels the job early without delays.}
    \label{figure:key-idea}
\end{figure}

\paragraph{\textbf{Dynamic Time Limit Adjustment.}}
The novel idea in this work is the dynamic adjustment of time limits for jobs that are already running. 
The concept is presented in Figure~\ref{figure:key-idea}. 
A misalignment of the user-provided time limit and the checkpoint schedule of an application leads to \textit{tail waste}. 
By canceling a job early or extending its time limit, we can align the timeout to the checkpoint schedule and minimize \textit{tail waste}. 
For this purpose, we define different policies for dynamic time limit adjustments:
\texttt{Early Cancellation} which cancels jobs after the last completed checkpoint (the last checkpoint that fits successfully during the initial time limit).
\texttt{Time Limit Extension} that always extends jobs to reach one more checkpoint regardless of other jobs in the queue (potentially delaying them).
\texttt{Hybrid Approach} which makes best effort decisions between \texttt{Early Cancellation} and \texttt{Time Limit Extension} without delaying subsequent jobs.

\paragraph{\textbf{Scheduling Metrics.}}
We rely on different scheduling metrics and characteristics, such as, job states (\texttt{COMPLETED} or \texttt{TIMEOUT}), makespan (the total time to complete a set of jobs), backfill statistics (either \texttt{SchedMain} or \texttt{SchedBackfill} reported by Slurm), and checkpoint counts (as reported by applications).
We also rely on total CPU time, \textit{tail waste} reduction, \textit{average job wait time} and \textit{weighted average job wait time}, which are defined below.

\paragraph{\textbf{CPU Time and Tail Waste.}}
CPU time is measured as the product of a job's execution time and the number of cores allocated to it. 
We define \textit{tail waste} as the CPU time spent on work that is lost because it occurs after the last completed checkpoint before the job timeout. 
Namely, \textit{tail waste} quantifies the unsaved computation that is not saved by a checkpoint. 
If a job terminates immediately after its last checkpoint, it has zero \textit{tail waste}. The total \textit{tail waste} of a workload is the sum of the \textit{tail waste} from each checkpointing job. In our context, non-checkpointing jobs do not have \textit{tail waste} even if they timeout.

\paragraph{\textbf{Average vs. Weighted Average Job Wait Time.}}
\textit{Average job wait time} is a common metric for evaluating job scheduling; 
however, optimizing for this metric has recently been shown to be misleading because it favors small jobs (which are traditionally the majority in HPC systems) at the expense of increasing waiting time of longer jobs~\cite{boezennec2024qualitatively,goponenko2022metrics}. 
As an alternative, we use the \textit{weighted average job wait time} where the job wait time is multiplied by a priority based on the allocated nodes to give a higher weight to larger jobs (measured in $nodes \times time$).

\section{Experimental Setup and Evaluation} \label{section:evaluation}

We evaluate the proposed time limit adjustment idea through implementation as an autonomy loop and experiments on a test system. This section offers details of the system on which we conducted the experiments and describes the experimental setup and workload used to evaluate our proposed approach.

\begin{figure}
    \centering
    \includegraphics[width=0.8\textwidth]{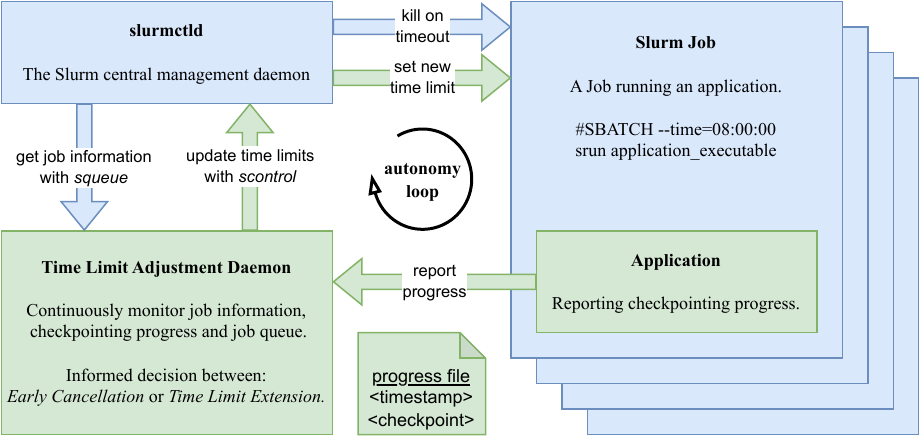}
    \caption{Autonomy loop architecture and interactions of the time limit adjustment daemon, the applications and the Slurm central management daemon \texttt{slurmctld}. A Slurm job executing an application reports its checkpointing progress through a temporary text file to the daemon; the daemon estimates the job's checkpointing interval, predicts the time of the next checkpoint, retrieves job queue information with \texttt{squeue}, decides whether early cancellation or time limit extension is appropriate, and issues update commands to \texttt{slurmctld} through \texttt{scontrol}. Slurm then sets the new time limits for the job as issued by the daemon.}
    \label{figure:daemon}
\end{figure}

\paragraph{\textbf{Daemon Architecture.}}
The time limit adjustment daemon is implemented in \texttt{Python} (v.~3.9.6), runs on the login node, and interacts with Slurm commands to adjust job time limits. As shown in Figure~\ref{figure:daemon}, checkpointing applications report checkpointing progress by writing timestamps to a temporary file; the daemon uses these to estimate the next checkpoint by adding the average checkpoint interval to the last checkpoint’s timestamp. 
Non-checkpointing jobs remain unchanged since they provide no progress information. The daemon polls the job queue (\texttt{squeue}) every 20 seconds to avoid overloading Slurm, tracks predicted start times and planned node allocations, and decides whether to use \textbf{\texttt{Early Cancellation}} or a \textbf{\texttt{Time Limit Extension}} via \texttt{scontrol}. This forms an autonomy loop among the application (reporting checkpoints), the daemon (calculating time-limit updates) and the Slurm central management daemon (\texttt{slurmctld}) (applying new limits). Checkpointing applications must include some mechanism for progress reporting; currently, we rely on writing a timestamp to a temporary file to mark each checkpoint.

\paragraph{\textbf{Experimental Setup.}}
Adjusting the time limits of executing jobs requires administrator access to Slurm (in particular for \texttt{scontrol}). 
This prerequisite limits our ability to conduct extensive testing in a production environment, since enabling a daemon to autonomously modify job configurations on a large-scale system is potentially disruptive.
However, existing Slurm simulators do not support this type of \textit{dynamic} adjustment to individual job configurations. 
Consequently, we conducted experiments on a smaller research cluster, where we could implement and evaluate our approach without disturbing production users. 
This involved applying filters to extract a smaller set of jobs from the available Marconi workload trace, scaling down the jobs in time, and adapting them (as synthetic dummy jobs) to match the available resources of our test environment.

\paragraph{\textbf{System Specifications.}}
Experiments were conducted on a University research cluster running \texttt{Slurm} (v.23.11) with a default scheduling configuration and using 20 \texttt{Intel Xeon E5-2640} nodes along with a login node and a storage node. 
Two networks support system operations: a 100 Gbit/s \texttt{Omni-Path} network for communication among compute nodes, and a 10 Gbit/s \texttt{Ethernet} network for user and administrator access. A two-level fat tree topology forms the basis of the \texttt{Omni-Path} interconnect.

\begin{figure}[!h]
    \centering
    \includegraphics[width=0.8\textwidth]{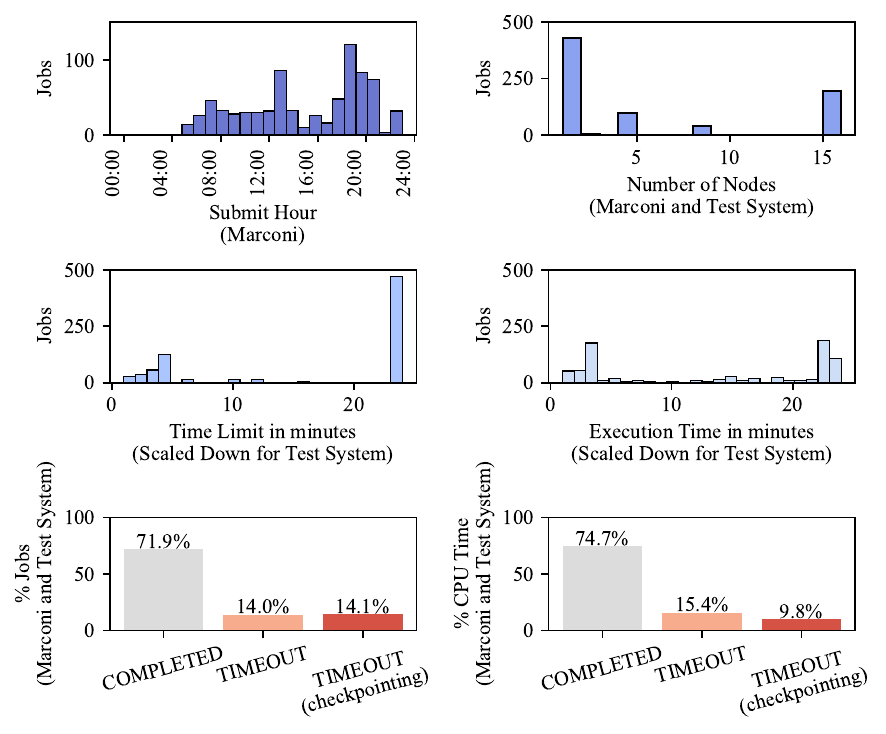}
    \caption{Overview of the 773 jobs selected and scaled down in time (reducing 1 hour to 1 minute) from the PM100 dataset.
    Shown are: the original submission time on Marconi, the original number of requested nodes, the scaled down user provided time limits, the scaled down execution time, the percentage of jobs by state, and the percentage of CPU time by state.}
    \label{figure:workload}
\end{figure}

\paragraph{\textbf{Workload Construction.}}
We derive our workload from the Marconi PM100 dataset of 1,074,576 jobs submitted between May and October 2020~\cite{antici2023pm100,andrea_borghesi_2023_7588815}.
We focused on the partition, queue, and month with most jobs (\texttt{Partition=1}, \texttt{Queue=1}, and \texttt{Month=May}). 
We filtered for jobs that were executed exclusively on their assigned nodes, with \texttt{COMPLETED} or \texttt{TIMEOUT} states, and ran for at least one hour, resulting in 773 jobs, shown in Figure~\ref{figure:workload}.

We adapted the Marconi jobs as synthetic sleep jobs with the same configurations; jobs that timeout at the maximum time limit of 24 hours (109 jobs) are adapted as checkpointing jobs that periodically report checkpoints at fixed-time intervals. 
To preserve the structure and dynamics of jobs while allowing the opportunity to study the impact of time limit adjustment on our smaller test system, we scaled job durations by a factor of 60 (one hour becomes one minute) and release all jobs at $t=0$. 
Jobs that were executed for less than an hour on Marconi were previously filtered out because after scaling down in time, they would only run for a few seconds, which is too short for meaningful experiments. 
The \texttt{checkpointing jobs} in our workload report successful checkpoints every 7 minutes to mimic a fixed-time interval checkpoint schedule that is misaligned with the job timeouts, though this interval can be adjusted to other values.

\section{Results} \label{section:results}

\begin{figure} [!h]
    \centering
    \includegraphics[width=0.8\textwidth]{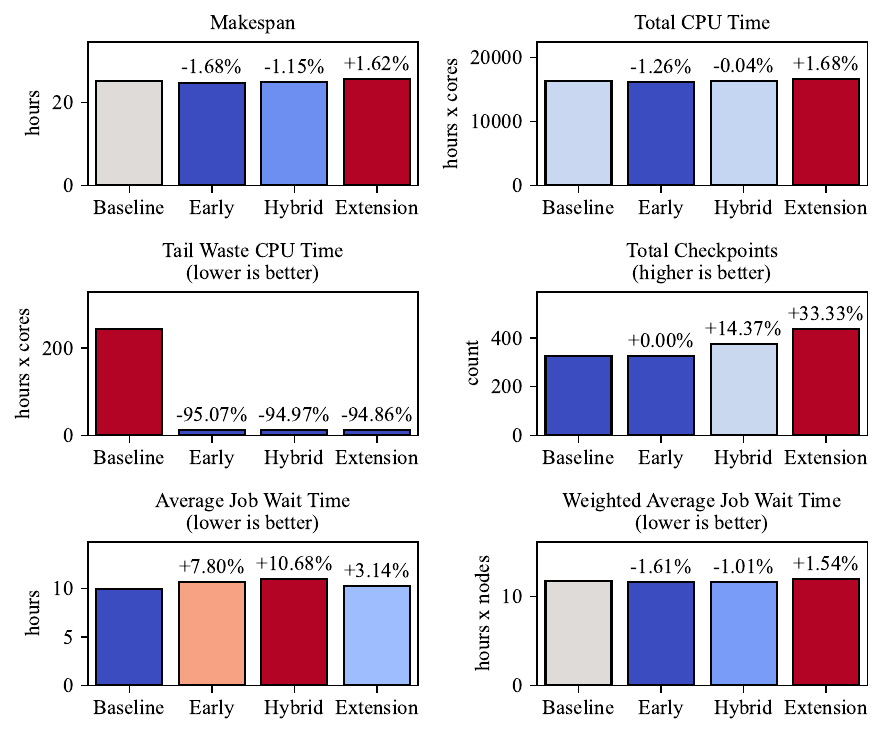}
    \caption{Comparison of scheduling metrics for the different time limit adjustment policies (\texttt{Early Cancellation}, \texttt{Time Limit Extension}, and \texttt{Hybrid} approach) against the \texttt{Baseline} (no time limit adjustment).}
    \label{figure:results}
\end{figure}

We present here the experimental results comparing the proposed job time limit adjustments against the baseline (no time limit adjustments). These are discussed later in Section~\ref{section:discussion}.
Table~\ref{table:results} summarizes the main findings and Figure~\ref{figure:results} illustrates key comparisons between the different policies.

\begin{table}[!h]
    \centering
    \caption{Comparison of scheduling scenarios under different daemon policies.}
    \begin{tabular}{l|rrrr}
        \toprule
        \textbf{Metric} (unit of measure) & \textbf{Baseline} & \textbf{Early} & \textbf{Time Limit} & \textbf{Hybrid} \\
        & & \textbf{Cancellation} & \textbf{Extension} & \textbf{Approach} \\
        \midrule
        \texttt{TIMEOUT} (jobs) & 217 & 108 & 108 & 108 \\
        \textbf{Early canceled (jobs)} & \textbf{--} & \textbf{109} & \textbf{--} & \textbf{62} \\
        \textbf{Extended time limit (jobs)} & \textbf{--} & \textbf{--} & \textbf{109} & \textbf{47} \\
        \texttt{COMPLETED} (jobs) & 556 & 556 & 556 & 556 \\
        Total Jobs (jobs) & 773 & 773 & 773 & 773 \\
        \midrule
        Slurm SchedMain (operations) & 203 & 189 & 202 & 201 \\
        Slurm SchedBackfill (operations) & 570 & 584 & 571 & 572 \\
        \midrule
        \textbf{Total Checkpoints} (count) & \textbf{327} & \textbf{327} & \textbf{436} & \textbf{374} \\
        \midrule
        Average Wait Time (sec) & 35,727 & 38,513 & 36,850 & 39,541 \\
        Weighted Avg Wait Time (nodes$\times$sec) & 42,349 & 41,666 & 43,001 & 41,923 \\
        \textbf{\textit{Tail Waste} CPU Time} (cores$\times$sec) & \textbf{875,520} & \textbf{43,120} & \textbf{45,020} & \textbf{44,000} \\
        Total CPU Time (cores$\times$sec) & 58,816,100 & 58,073,280 & 59,804,280 & 58,795,320 \\
        \midrule
        Workload Makespan (sec) & 90,948 & 89,424 & 92,420 & 89,901 \\
        \bottomrule
    \end{tabular}
    \label{table:results}
\end{table}

\paragraph{\textbf{Job Outcomes and Checkpointing.}}
Each policy in Table~\ref{table:results} runs the same 773 jobs: 556 always \texttt{COMPLETED} and 217 \texttt{TIMEOUT} under the \textit{Baseline}. Of these 217, 109 are assumed to use checkpointing (they timeout at the maximum time limit on Marconi), while 108 are assumed to be non-checkpointing and remain \texttt{TIMEOUT} under all new policies. \texttt{Early Cancellation} immediately cancels those 109 checkpointing jobs after their last completed checkpoint; \texttt{Time Limit Extension} extends them to the next checkpoint; and \texttt{Hybrid} cancels 62 and extends 47 jobs. Consequently, \textit{Baseline} and \textit{Early Cancellation} both achieve 327 successful checkpoints, while \textit{Hybrid} yields 374, and \textit{Time Limit Extension} reaches 436.

\paragraph{\textbf{Tail Waste Reduction.}}
In our experiments (Table~\ref{table:results}, Figure~\ref{figure:results}), the baseline scenario incurred 875,520 ($cores \times seconds$) of tail waste. 
In comparison: \texttt{Early Cancellation} cut tail waste to 43,120 ($cores \times seconds$), a 95.1\% reduction, also saving approximately 1.3\% of total CPU time by canceling jobs immediately after their last completed checkpoint before the timeout. 
\texttt{Time Limit Extension} achieved a 94.8\% reduction, replacing what would have been wasted CPU time with additional \textit{useful} work through an additional checkpoint. 
\texttt{Hybrid} reduced tail waste by 95\% by combining early cancellation with selective time limit extensions (only extending if it does not delay subsequent jobs). 
All three policies cut tail waste by roughly 95\%. 
However, only \texttt{Early Cancellation} explicitly shows the net 1.3\% CPU time saving; \texttt{Time Limit Extension} and \texttt{Hybrid} also save the same wasted time but use it for extra checkpoints, thus increasing the total CPU time in the results.

\paragraph{\textbf{Makespan and CPU Time.}}
Compared to the baseline, \texttt{Early Cancellation} reduces the makespan by 1.7\%, while \texttt{Hybrid} approach reduces it by 1.2\%, shown in Figure~\ref{figure:results}. Since \texttt{Time Limit Extension} always extends all jobs to complete another checkpoint, it increases the makespan by 1.6\% (of useful work because it was saved through a checkpoint). \texttt{Early Cancellation} reduces total CPU time by 1.3\% (total CPU time saved by reducing \textit{tail waste}). \texttt{Hybrid} approach remains close to the baseline, and \texttt{Time Limit Extension} increases CPU time by 1.7\%.

\paragraph{\textbf{Job Wait Times and Backfilling.}}
The \textit{average job wait time} increases by 7.8\% for \texttt{Early Cancellation}, 10.7\% for \texttt{Hybrid} approach, and 3.14\% for \texttt{Time Limit Extension}.
However, the \textit{weighted average job wait time} decreases by -1.6\% for \texttt{Early Cancellation} and -1.0\% for \texttt{Hybrid} approach, and increases by +1.5\% for \texttt{Time Limit Extension}, as shown in Figure~\ref{figure:results}. 
The only notable difference in the backfill statistics is in the \texttt{Early Cancellation} policy, showing a slight increase of 10 jobs started by Slurm's backfill scheduler \texttt{SchedBackfill} instead of the priority scheduler \texttt{SchedMain}, as shown in Table~\ref{table:results}.

\section{Discussion} \label{section:discussion}
The \texttt{Early Cancellation}, \texttt{Time Limit Extension}, and \texttt{Hybrid} policies achieve the goal of reducing \textit{tail waste}, by approximately 95\%, saving approximately 1.3\% of total CPU time. Compared to the \texttt{Baseline}, with misaligned timeouts, \texttt{Early Cancellation} achieves the highest reduction in total CPU time and makespan by proactively freeing resources earlier, while keeping the same number of checkpoints. \texttt{Time Limit Extension} achieves more checkpoints, but does so at the cost of increasing total CPU time and makespan because it extends jobs to accommodate an additional checkpoint; it is suited for environments that prioritize additional \textit{useful} work even if others jobs are delayed. The \texttt{Hybrid} approach combines both strategies, freeing resources early when another checkpoint does not fit within the original time limit and extending time limits only if it does not delay other jobs.

\paragraph{\textbf{Benefits and Trade-Offs.}}
The proposed policies significantly reduce wasted computing time. 
These benefits scale with the proportion of jobs that use checkpoints, more checkpointing jobs translate into more potentially saved \textit{tail waste}. 
However, policies for extending jobs must be carefully calibrated to avoid delaying other queued jobs. 
For large-scale HPC centers, \texttt{Early Cancellation} and \texttt{Hybrid} are more attractive because they reduce \textit{tail waste} and preserve or improve scheduling performance without delaying other jobs. 
\texttt{Time Limit Extension} is preferred where every possible checkpoint is important and the increase of total CPU time and makespan is acceptable. 

It is important to note that the higher total CPU time under \texttt{Hybrid} and \texttt{Time Limit Extension} reflects additional \textit{useful} work via successful checkpoints rather than wasted resources. Although \textit{average job wait time} rises for all policies, largely because this metric is skewed by many small jobs, \textit{weighted average job wait time} decreases with \texttt{Early Cancellation} and \texttt{Hybrid}, denoting a more efficient use of resources and job \textit{scheduling}, achieved by freeing resources earlier and only extending the time limits of running jobs if they do not delay other jobs. In contrast, \texttt{Time Limit Extension} increases \textit{weighted average job wait time} because it always extends the running jobs and potentially delays other jobs in the queue.

\paragraph{\textbf{Impact on Scheduling with Backfilling.}}
Only \texttt{Early Cancellation} has a significant impact on whether a job was started by the \texttt{SchedMain} or \texttt{Schedbackfill} scheduler, by slightly increasing the number of backfilled jobs by 10. 
This is explained by the increased \textit{free resource windows} created by \texttt{Early Cancellation} (which frees resources earlier than expected at the start of the job), allowing other jobs to be started sooner by the backfill scheduler. 
No such impact is observable with the other policies.
This is also explained by the fact that \texttt{Hybrid} approach and \texttt{Time Limit Extension} do not always cancel jobs early, but also or always extend their time limits, possibly extending an already running job into an empty resource window that would otherwise be used for backfilling.

\paragraph{\textbf{Limitations.}}
A limitation of the proposed approach is the reliance on accurate application checkpoint reporting. 
If jobs delay or misreport their checkpointing progress, or if there is strong variation among the checkpoint intervals, the daemon's prediction of the next checkpointing time may be inaccurate. 
Additionally, the \texttt{Time Limit Extension} policy, if overused, can negatively impact scheduling by delaying subsequent jobs. 
It is also an open question whether users and application developers will adopt this approach by reporting checkpoint progress.
However, implementing checkpoint reporting (printing a timestamp into a temporary file after a checkpoint) is relatively easier compared to implementing a dynamic checkpointing approach which involves potential code restructuring. 
This can drive adoption, since HPC users prefer simplicity over complexity in the context of checkpointing~\cite{bautista2024survey,el2016understanding}.

\section{Related Work} \label{section:related-work}

Changing the job scheduler has been the focus of recent work, which proposed adapting scheduling decisions based on real-time monitoring of I/O usage~\cite{goponenko2020towards} to avoid file system bottlenecks.
This approach integrates \texttt{LDMS} (Lightweight Distributed Metric Service)~\cite{agelastos2014lightweight} and Slurm to capture real-time \texttt{Lustre} throughput and job-level I/O usage.
Another related approach is to leverage deep reinforcement learning (DRL) for the prediction of the remaining execution time~\cite{wang2021rlschert}.
Using a neural network to estimate the remaining execution time of each job by analyzing intermediate logs, it employs a DRL policy to decide which jobs to schedule, preempt, or kill. 
Other work focused on predicting failures and triggering early termination~\cite{zasadzinski2018early}, leveraging a neural network trained on system metrics and job history to predict job evolution and states.
The use of reinforcement learning to dynamically schedule jobs has also recently been proposed~\cite{zhang2020rlscheduler}. 
The approach relies only on minimal manual interventions or expert knowledge; it can continuously learn and improve scheduling decisions and policies over time and uses a kernel-based neural network.
Our proposed approach differs from the above approaches in that we do not directly change the configuration or decision making of the job scheduler. 
Rather, we indirectly change the scheduling of jobs by adjusting the time limits of jobs that are \textit{already} running.

Another line of research involves predicting job time limits to provide better accuracy for backfilling.
A hierarchical classification scheme has been proposed to improve the user-provided estimates~\cite{lamar2021backfilling}. 
This approach avoids underestimation by making a prediction that is higher than a specified percentage of the observed job execution times.
Another approach focused on the prediction of time limits~\cite{tsafrir2007backfilling} separated the kill time from the time limit prediction to avoid jobs being killed because system predictions were wrong. 
The predictions were achieved by averaging the execution times of the last two jobs by the same user.
In contrast to these efforts, which make predictions about job execution times, we focus on the novel perspective of adjusting the time limits of \textit{already} running jobs. 
Our approach exploits knowledge about the progress of the application and the workload in the job queue, which becomes available only at execution time.
We do not change configurations of jobs waiting in the queue, but directly incorporate progress information and change time limits of jobs that are already executing.

\section{Conclusion and Future Work} \label{section:conclusion}

This work proposed a mechanism for implementing an autonomy loop to dynamically adjust HPC job time limits based on their checkpointing progress. 
By analyzing checkpointing intervals and queued jobs, a time limit adjustment daemon decides whether to extend time limits or gracefully end a job, minimizing wasted computations. 
Through experimental results, we showed that this adaptive policy avoids resource waste due to timeouts by reducing \textit{tail waste} (i.e., the unsaved computations made between the last completed checkpoint and the timeout) by approximately 95\% and saving approximately 1.3\% of total CPU time that would otherwise be wasted.
The proposed solution shows that real-time application-specific information can be leveraged via an autonomous loop to reduce resource waste and improve job scheduling performance.
As HPC systems and workloads continue to grow, fully automated feedback mechanisms such as the one proposed in this work will become increasingly essential to manage workload complexity and optimize overall system performance.

\textit{Future work} can integrate real-time I/O load to account for the potential slowdown of checkpoints due to system noise; fine-tune checkpoint predictions based on historical/other data from the respective applications; and perform larger experiments with real application workloads in a production(-like) system.

\bibliographystyle{splncs04}
\bibliography{mybibliography.bib}

\end{document}